\documentclass[aip,rsi,amsmath,amssymb,reprint,]{revtex4-1}

\usepackage{graphicx}
\usepackage{dcolumn}
\usepackage{bm}
\usepackage[utf8]{inputenc}
\usepackage[T1]{fontenc}
\usepackage{mathptmx}

\usepackage{amsmath}
\usepackage{amssymb}
\usepackage{color}
\usepackage{multirow}

\newcommand{\II}{{\mathbb{I}}}

\begin{document}


\title{Laser control of molecular rotation:\\ Expanding the utility of an optical centrifuge}

\author{Ian~MacPhail-Bartley}
\author{Walter~W.~Wasserman}
\author{Alexander~A.~Milner}
\author{V.~ Milner}
\email{vmilner@phas.ubc.ca}

\affiliation{$^{1}$Department of  Physics \& Astronomy, The University of British Columbia, Vancouver V6T-1Z1, Canada}

\date{\today}

\begin{abstract}
Since its invention in 1999, optical centrifuge has become a powerful tool for controlling molecular rotation and studying molecular dynamics and molecular properties at extreme levels of rotational excitation. The technique has been applied to a variety of molecular species, from simple linear molecules to symmetric and asymmetric tops, to molecular ions and chiral enantiomers. Properties of isolated ultrafast rotating molecules, so-called molecular superrotors, have been investigated, as well as their collisions with one another and interaction with external fields. The ability of an optical centrifuge to spin a particular molecule of interest depends on both the molecular structure and the parameters of the centrifuge laser pulse. An interplay between these two factors dictates the utility of an optical centrifuge in any specific application. Here, we discuss the strategy of assessing and adjusting the properties of the centrifuge to those of the molecular rotors, and describe two practical examples of optical centrifuges with very different characteristics, implemented experimentally in our laboratory.
\end{abstract}

\maketitle

\section{Introduction}
\label{sec:Introduction}
\begin{figure}[b]
    \centering
    \includegraphics[width=.75\columnwidth]{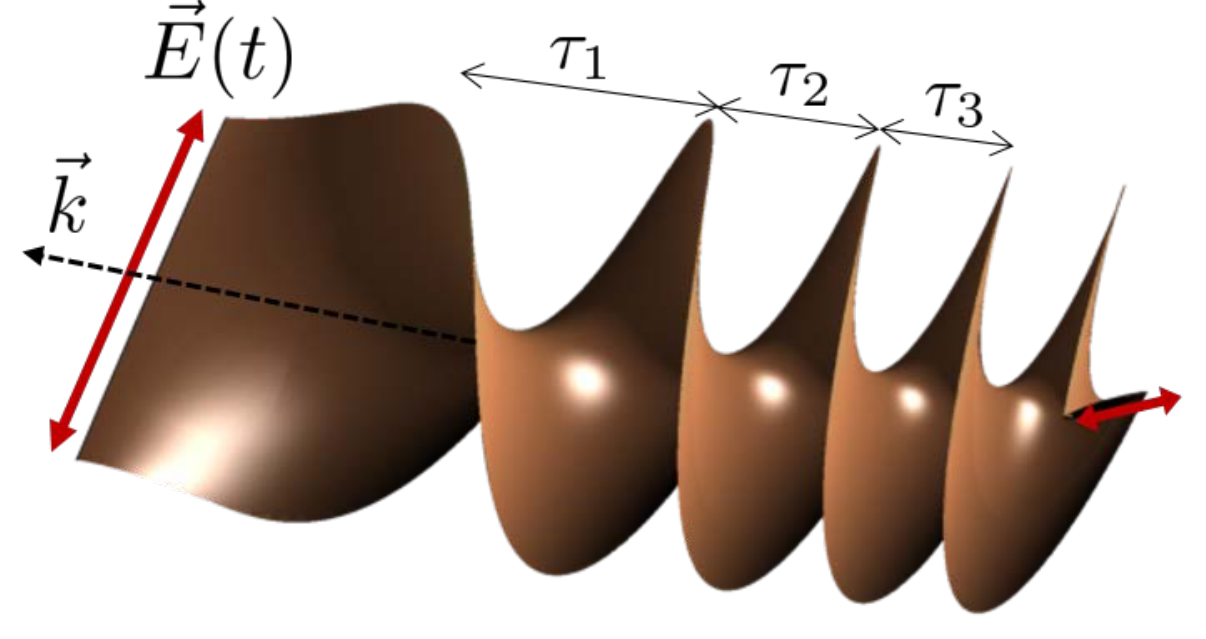}
    \caption{The field of an optical centrifuge, propagating in the direction of wave vector $\vec{k}$. The vector of linear polarization $\vec{E}$ follows the surface of the corkscrew shape. As the pitch of the corkscrew becomes shorter along the centrifuge pulse, i.e. $\tau _1>\tau _2>\tau _3$, the instantaneous angular frequency of the rotating polarization grows with time.}
    \label{fig:CFGScrew}
\end{figure}

Intense laser pulses have been long used for controlling rotation and spatial orientation of molecules (for reviews on this broad topic, see Refs.~\citenum{Stapelfeldt2003,Ohshima2010,Fleischer2012}). The control mechanism is based on the interaction of the electric field of an optical wave with the induced electric dipole of the molecule. Among multiple approaches to the rotational control, the method of an optical centrifuge was proposed in 1999 by Karczmarek \textit{et al.} \cite{Karczmarek1999} It involved a linearly polarized laser pulse, whose polarization vector rotates with a constant angular acceleration, as illustrated in Figure~\ref{fig:CFGScrew}. The field of the centrifuge induces an oscillating dipole moment, whose direction is determined by the polarizability tensor of the molecule and the orientation of its axes with respect to the field vector. An interaction of the induced dipole moment with the applied laser field results in a torque, which pushes the most polarizable molecular axis towards the field polarization. If the latter is rotating around the propagation direction of the laser beam, the same torque will force the molecule to follow this rotation, much like an object placed inside a mechanical centrifuge follows its rotary motion.

In comparison to other schemes of rotational excitation where a molecule interacts with an ultrashort laser pulse, an optical centrifuge extends the reach of rotational control in two important ways: it offers the directionality of molecular rotation (i.e. dictates whether the molecules rotate clockwise or counterclockwise with respect to a laboratory fixed axis), and it enables one to spin the molecules to the desired angular frequency. The first aspect is not unique to the centrifuge, as unidirectional molecular rotation has also been generated with a sequence of short laser pulses \cite{Fleischer2009,Kitano2009,Zhdanovich2011,Karras2015}. On the other hand, the ability to create a narrow rotational wave packet centered at a well-defined target rotational state is a distinctive feature of the centrifuge. Moreover, the target state can be controlled in a broad range of angular momenta, and corresponding rotational frequencies, which are inaccessible by other means such as a single pulse or a train of pulses \cite{Milner2016a}.

As a powerful tool for controlling molecular rotation, optical centrifuges have been used in multiple experimental studies of molecular structure and internal properties of neutral molecules \cite{Villeneuve2000,Yuan2011a,Korobenko2014b,Milner2014b,Milner2017b} and molecular ions \cite{Vestergaard2020}, molecular collisions \cite{Yuan2011,Toro2013,Korobenko2014a,Milner2014a,Murray2018}, dynamics \cite{Korobenko2014c,Milner2015c} and alignment \cite{Korobenko2016a,Milner2016b}, molecular interactions with external fields \cite{Korobenko2015a,Milner2017a}, the study of molecular chirality \cite{Milner2019} and control of chemical reactions \cite{Ogden2019}. It has also been found that molecules spun by the centrifuge, known as ``superrotors'', exhibit unique optical and magnetic properties \cite{Milner2014c,Faucher2016,Floss2018}.

The two most important parameters of an optical centrifuge, related to its ability to spin molecules, are the intensity $I$ and the angular acceleration $\beta $. The former defines the well depth of a rotating dipole trap, whereas the latter describes its dynamics. Both parameters depend on the laser source -- its pulse energy and spectral bandwidth -- as well as on the optical setup used for shaping the pulse into the polarization corkscrew of the centrifuge. In this report we discuss two operational limits of an optical centrifuge. On one side, pulses with a relatively high frequency bandwidth of 20~THz (30~fs Fourier transform limit, TL) offer high terminal rotational frequencies which can be reached in about 100~ps. On the other side, longer 120~fs pulses (5~THz bandwidth) enable longer angular acceleration on the time scale of 500~ps to lower terminal frequencies. The former limit, hereafter referred to as ``fast centrifuge'' (fCFG), provides faster rotating molecules, whereas the latter (``slow centrifuge'', sCFG) offers better adiabaticity (as will be discussed later). Fast centrifuge is well suited for the studies of extreme rotational states. Slow centrifuge, on the other hand, can be advantageous for spinning molecules with higher moment of inertia or molecules whose rotation is impeded by the surrounding viscous medium (e.g. molecules embedded in helium nanodroplets \cite{Toennies2004}). Both slow and fast centrifuges have been built in our laboratory. In what follows, we describe the details of their construction and demonstrate their properties.

In Section~\ref{sec:Theory}, we will assess the efficiency of an optical centrifuge in two temperature regimes, and discuss its dependence on the properties of the molecule and on the parameters of the centrifuge. In Section~\ref{sec:MakingCFG}, technical aspects and constraints of building an optical centrifuge in the limit of either fast or slow angular acceleration will be presented. Section~\ref{sec:Characterization} will be devoted to the methods of characterizing centrifuge pulses. Two representative examples of applying both fast and slow centrifuges to the gas of oxygen and nitrogen molecules at room temperature will be shown and used to demonstrate the difference in the two regimes of rotational control. Section~\ref{sec:Conclusion} will summarize the main results of our study.

\section{Molecular Spinning with an Optical Centrifuge}
\label{sec:Theory}
The non-resonant interaction of a symmetric top molecule (taken as a simpler example) with a linearly polarized optical field $\vec{E}=\hat{e} E_0 \cos(\omega_0 t)$ can be described by the following potential averaged over the period of carrier oscillations ($2\pi/\omega_0$):
\begin{equation}\label{eq:UThetaU0}
U(\theta) = -U_0 \cos^2(\theta), \hspace{0.5cm}
U_0 = \frac{1}{4} E_0^2 \Delta \alpha,
\end{equation}
\noindent where $\theta$ is the angle between the polarization vector $\hat{e}$ and the most polarizable molecular axis, and the polarizability anisotropy $\Delta \alpha = \alpha_\parallel - \alpha_\perp $ is the difference of the polarizability components along ($\alpha_\parallel$) and perpendicular ($\alpha_\perp$) to the molecular symmetry axis. $E_0$ is a slowly varying envelope, whose dependence on time is omitted for clarity. In the simplest case of a rigid rotor, potential~(\ref{eq:UThetaU0}) exerts an angle-dependent torque $\tau = \partial U/\partial \theta  = U_0\sin(2\theta)$. This torque is forcing the molecule to rotate towards the bottom of the potential well at $\theta = (0,\pi)$ with an average angular acceleration
\begin{equation}\label{eq:EpsilonMol}
\epsilon_{\text{mol}}  = \frac{\tau}{\II} \approx \frac{2 U_0}{\pi \II},
\end{equation}
\noindent where $\II$ is the molecular moment of inertia. If the polarization vector $\hat{e}$ (and with it, the axis of the potential well) changes its direction with acceleration $\epsilon_{\text{pol}} $, the molecule will follow the field adiabatically as long as $\epsilon_{\text{pol}} < \epsilon_{\text{mol}}$, or
\begin{equation}\label{eq:Adiabaticity}
\epsilon_{\text{pol}} < \frac{2 U_0}{\pi \II}.
\end{equation}

Following Karczmarek \textit{et al.} \cite{Karczmarek1999}, consider the interference of two laser pulses of opposite circular polarizations ($\hat{e}_\pm = \hat{x} \pm i\hat{y}$) and carrier frequencies shifted by $\pm\Omega$ from $\omega _0$, respectively, propagating along $\hat{z}$:
\begin{eqnarray}\label{eq:EplusEminus}
\vec{E}_{+}(t)=\frac{E_0}{2}\left[\hat{x}\cos(\omega_0 + \Omega) t + \hat{y}\sin(\omega_0 + \Omega) t \right], \\
\vec{E}_{-}(t)=\frac{E_0}{2}\left[\hat{x}\cos(\omega_0 - \Omega) t - \hat{y}\sin(\omega_0 - \Omega) t \right]. \nonumber
\end{eqnarray}
\noindent The result of such interference is an optical field, whose linear polarization is rotating in the $xy$ plane with a constant angular frequency $\Omega$,
\begin{equation}\label{eq:Ecircular}
    \vec{E}_\circlearrowleft(t)=\vec{E}_{+}(t)+\vec{E}_{-}(t)= E_0\cos(\omega_0 t)\left[\hat{x} \cos(\Omega t) + \hat{y} \sin(\Omega t) \right].
\end{equation}
\noindent To make the polarization vector rotate with a gradually increasing angular speed ($\Omega = \epsilon_{\text{pol}} t$), the frequency difference between the two interfering fields must be increasing with time. This could be achieved by chirping the frequencies of the two pulses in opposite directions, i.e. by adding quadratic phase factors of opposite sign,
\begin{equation}\label{eq:Eplusminus}
\vec{E}_{\pm}(t)=\frac{E_0}{2}\left[\hat{x}\cos(\omega_0 t + \beta_\pm t^2) \pm \hat{y}\sin(\omega_0 t + \beta_\pm t^2) \right],
\end{equation}
\noindent with $\beta_\pm \gtrless 0$, and the average chirp rate $\beta \equiv (\beta _+ - \beta _-)/2 >0$ defining the instantaneous angular frequency and acceleration as follows:
\begin{equation}\label{eq:EpsilonPpol}
\Omega(t) = 2\beta t, \hspace{0.5cm} \epsilon_{\text{pol}} = \dot{\Omega }(t) = 2\beta.
\end{equation}
\begin{figure*}[t]
    \centering
    \includegraphics[width=.8\textwidth]{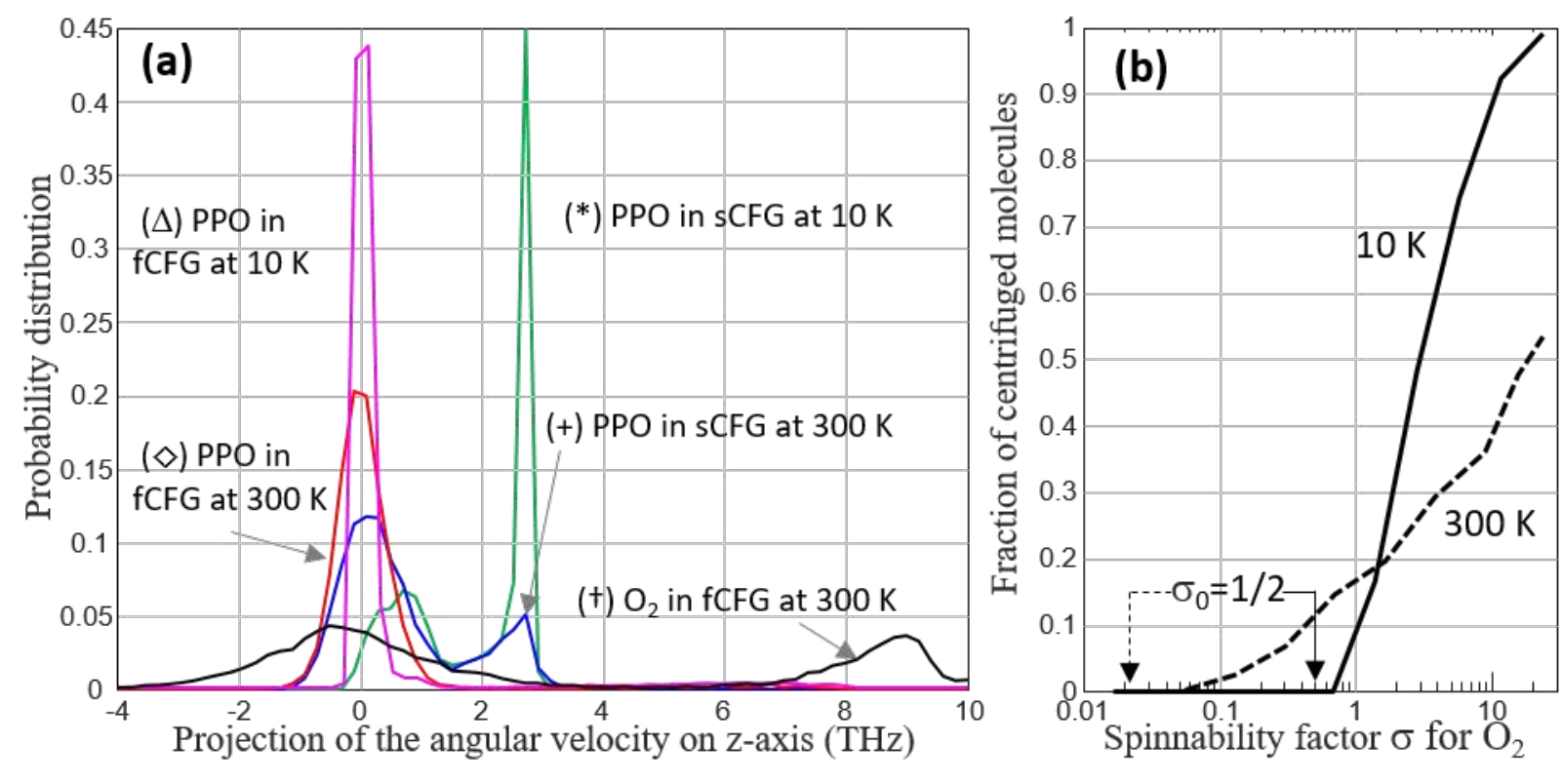}
    \caption{\textbf{(a)} Numerically calculated probability distributions of the projection of the molecular angular velocity on the propagation direction of the centrifuge. Different curves correspond to the five cases labeled with superscripts `$*,+,\vartriangle,\diamond$' and `$\dag$' in Table~\ref{tabl:Spinnability}; the curves are labeled with the same symbols. \textbf{(b)} Numerically calculated total fraction of centrifuged oxygen molecules as a function of the spinnability factor $\sigma$ for the initial gas temperature of 10~K (solid line) and 300~K (dashed line). Solid and dashed arrows mark the threshold value of $\sigma _0= 1/2$ for the low and high temperature, respectively.}
    \label{fig:Spinnability}
\end{figure*}

Substituting the acceleration of the centrifuge into Eq.~(\ref{eq:Adiabaticity}) leads to the following adiabaticity condition:
\begin{equation}\label{eq:Sigma0}
\sigma_0 := \frac{U_0}{\pi \II \beta} = \frac{1}{4\pi} \times \frac{E^2_0}{\beta} \times \frac{\Delta \alpha }{\II} > 1,
\end{equation}
\noindent where factors $E^2_0/\beta$ and $\Delta \alpha /\II$ characterize the field and the molecule, respectively. For a rotationally cold molecule, Equation~(\ref{eq:Sigma0}) guarantees that the molecule will be trapped in a rotating potential well \cite{Karczmarek1999}. This is true when the initial energy of rotation around the relevant molecular axis ($\tfrac{1}{2} kT$, where $k$ is the Boltzmann constant and $T$ the temperature) is lower than the kinetic energy acquired by the molecule during the turn-on time of the centrifuge $t_{\text{on}}$, $U_0/2+\II(2\beta t_{\text{on}})^2/2$.

In the opposite limit of high initial temperature, one can assess the overall spinning efficiency by assuming that the centrifuge should slow the molecule down before forcing it to follow the rotating polarization vector. This requirement can be expressed as $\beta t^2_{\text{stop}} < \pi/2$, where $t_{\text{stop}}$ is the characteristic slow-down time. The latter can be estimated as the time needed for decelerating the molecule from the most probable angular frequency $\sqrt{kT/\II}$ down to zero with an average acceleration $\epsilon_{\text{mol}}$. Using Eq.~(\ref{eq:EpsilonMol}) we obtain $t_{\text{stop}}=\pi\sqrt{kT\II}/(2U_0)$ and an additional trapping criterion:
\begin{equation}\label{eq:SigmaT}
\sigma_T:=\frac{U_0}{\pi \II \beta}\frac{U_0}{kT/2} =\sigma_0 \times \frac{U_0}{kT/2} >1,
\end{equation}
\noindent which \textit{together with} $\sigma_0>1$ would define the conditions for efficient rotational excitation of initially hot molecules. Recent theoretical analysis by Armon \textit{et al.} \cite{Armon2016} showed that the trapping mechanism is more complex than the simplified picture described above. As long as $\sigma_0>1/2$, the centrifuge may capture molecules while sweeping through (rather than slowing down) the initially untrapped thermal distribution even in the weak-field limit of $U_0 \ll kT/2$.

\begin{table}[b]
\centering
\begin{tabular}{|l|r|r|r|r|}
\hline
\textbf{Molecule}       & \multicolumn{2}{l|}{\textbf{Slow Centrifuge}} & \multicolumn{2}{l|}{\textbf{Fast Centrifuge}}         \\ \cline{2-5}
($\Delta\alpha/\II$)    & $T=10$~K              & $T=300$~K                 & $T=10$~K                          & $T=300$~K                     \\ \hline
Oxygen                  & \multirow{2}*{106}    & \multirow{2}*{57}         & \multirow{2}*{6.2}                & \multirow{2}*{3.3$^\dag$}     \\
($6.1\times 10^5$)      &                       &                           &                                   &                               \\ \hline
Nitrogen                & \multirow{2}*{93}     & \multirow{2}*{32}         & \multirow{2}*{5.4}                & \multirow{2}*{1.9}            \\
($5.4\times 10^5$)      &                       &                           &                                   &                               \\ \hline
Propylene oxide         & \multirow{2}*{16$^*$} & \multirow{2}*{8.8$^+$}    & \multirow{2}*{0.9$^\vartriangle$} & \multirow{2}*{0.5$^\diamond$} \\
($0.9\times 10^5$)      &                       &                           &                                   &                               \\ \hline
\end{tabular}
\caption{Calculated spinnability $\sigma $ [Eq.~(\ref{eq:Sigma})] at $I=5\times10^{12}$~W/cm$^{2}$ and various experimental conditions. Values of $\Delta\alpha/\II$ in brackets are in SI units of V$^{-2}$m$^2$s$^{-2}$. Superscripts `$*,+,\vartriangle,\diamond$' and `$\dag$' indicate five limiting cases, numerically simulated in Figure~\ref{fig:Spinnability}\textbf{(a)}.}
\label{tabl:Spinnability}
\end{table}

Assuming for simplicity that the centrifuge efficiency is governed predominantly by either $\sigma _0$ or $\sigma _T$, whichever is smaller, let us estimate the value of the ``spinnability'' factor
\begin{equation}\label{eq:Sigma}
\sigma := min\{\sigma _0, \sigma _T\}
\end{equation}
\noindent for a few molecular species and a number of realistic experimental conditions. We take the field strength $E_0$ corresponding to the laser intensity of $5\times 10^{12}$~W/cm$^2$, typically dictated by the onset of either strong photo-ionization and/or filamentation, both at $I\approx 10^{13}$~W/cm$^2$. For the frequency chirp $\beta $, we consider the two practical limits of $0.3$~rad/ps$^{2}$ and $0.017$~rad/ps$^{2}$ for, respectively, the fast and slow optical centrifuges described in Section~\ref{sec:Introduction} (the exact numerical values are explained in Sections~\ref{sec:MakingCFG} and \ref{sec:Characterization} below). We obtain molecular polarizabilities and moments of inertia from the NIST database (method B3LYP/cc-pVTZ) \cite{NIST}. Finally, two temperature limits are considered, $T=10$~K and $T=300$~K, characteristic of a typical molecular beam expansion and room temperature ensembles, respectively. The results are shown in Table~\ref{tabl:Spinnability}. One can see that the spinnability factor decreases with both the temperature and the speed of the centrifuge increasing (left to right) as well as with the decreasing ratio $\Delta\alpha/\II$ specific to each particular molecule (top to bottom). This demonstrates the advantage of lowering the acceleration of the centrifuge for implementing rotational control at higher temperatures and/or applying it to more complex molecular systems.

To illustrate the practical meaning of $\sigma $ [Eq.~(\ref{eq:Sigma})], we performed classical simulations of the centrifuge action on the molecules of oxygen (O$_{2}$) and propylene oxide (PPO, CH$_3$CHCH$_2$O) under experimental conditions (temperature and frequency chirp) discussed above and labeled with superscripts `$*,+,\vartriangle,\diamond$' and `$\dag$' in Table~\ref{tabl:Spinnability}. The calculation procedure is based on expressing both the orientation and the angular velocity of a molecule (in the rigid body approximation) by means of quaternions \cite{Tutunnikov2018}. The coupled system of Euler equations and quaternion equations of motion is then solved numerically. For the initial conditions, we took a thermal ensemble of 10,000 molecules and assumed an isotropic distribution of molecular axes and Maxwell-Boltzmann distribution of angular velocities. Peak intensity of the centrifuge field was set at $5\times 10^{12}$~W/cm$^{2}$ for the reason discussed earlier in the text. Since the centrifuge affects the projection of the molecular angular velocity along the laser propagation direction, we calculated this quantity for each molecule in the ensemble. The final probability distributions of such projections are shown in Figure~\ref{fig:Spinnability}\textbf{(a)}. In each case, the relative amount of centrifuged molecules can be found by integrating the corresponding distribution between 5 and 10~THz for the fast centrifuge, or between 1.5 and 4~THz for the slow one.

On the lowest side of the considered spinnability range is the case of a room temperature gas of PPO molecules in the fast centrifuge ($\sigma=0.5$ in Table~\ref{tabl:Spinnability}). Here, only about 2\% of molecules are caught by the centrifuge, with all of them falling out of the rotating trap prematurely, i.e. without reaching the terminal frequency of fCFG at about 9~THz. This is reflected by the broad and low (barely visible) shoulder of the red curve labeled with `$\diamond$'. In contrast, applying an identical centrifuge to oxygen -- a molecule with a higher $\Delta\alpha/\II$ ratio, results in a centrifuged fraction of $\approx30\%$ (black curve labeled with `$\dag$'). Improving the spinnability of propylene oxide by lowering the temperature of the gas is not very efficient. This is demonstrated by the magenta curve (`$\vartriangle$', $\sigma=0.9$ in Table~\ref{tabl:Spinnability}). Apparently, the centrifuge accelerates too fast for catching even cold PPO molecules. On the other hand, much higher spinning efficiency can be achieved by using a slow centrifuge. This is demonstrated by the blue curve labeled with `+', which shows that more than 20\% of PPO molecules have followed sCFG to its terminal frequency of 2.5~THz at room temperature. Ultimately, applying the same slow centrifuge to a cold gas brings the fraction of PPO superrotors to 65\% (green curve marked with `*').

As shown in Refs.~\citenum{Armon2016, Armon2017}, the efficiency of the centrifuge cannot be parametrized by a single parameter. Indeed, when we plot the numerically calculated relative amount of centrifuged molecules as a function of the spinnability factor $\sigma $ introduced here, the results do not obey a single scaling law. This is illustrated in Fig.~\ref{fig:Spinnability}\textbf{(b)} for the case of oxygen gas at 10~K (solid line) and 300~K (dashed line). In both cases, the centrifuge action begins at $\sigma _0  > 1/2$ (marked with solid and dashed arrows, respectively) \cite{Armon2016}. In the low-temperature case, the centrifuge is strong enough to trap the majority of molecules in a thermal ensemble, and its efficiency is determined by the adiabaticity of spinning. The high-temperature case, on the other hand, belongs to the weak-field regime of interaction, where the trapping capability governs the outcome of the centrifuge action. Despite the lack of universality, the two curves in Figure~\ref{fig:Spinnability}\textbf{(b)} demonstrate the general dependence of the centrifuge efficiency on $\sigma$ in two temperature limits, and the amount by which it can be improved with lower angular acceleration $\beta $ ($\sigma \propto \beta ^{-1}$).
\begin{figure*}[t]
    \centering
    \includegraphics[width=.7\textwidth]{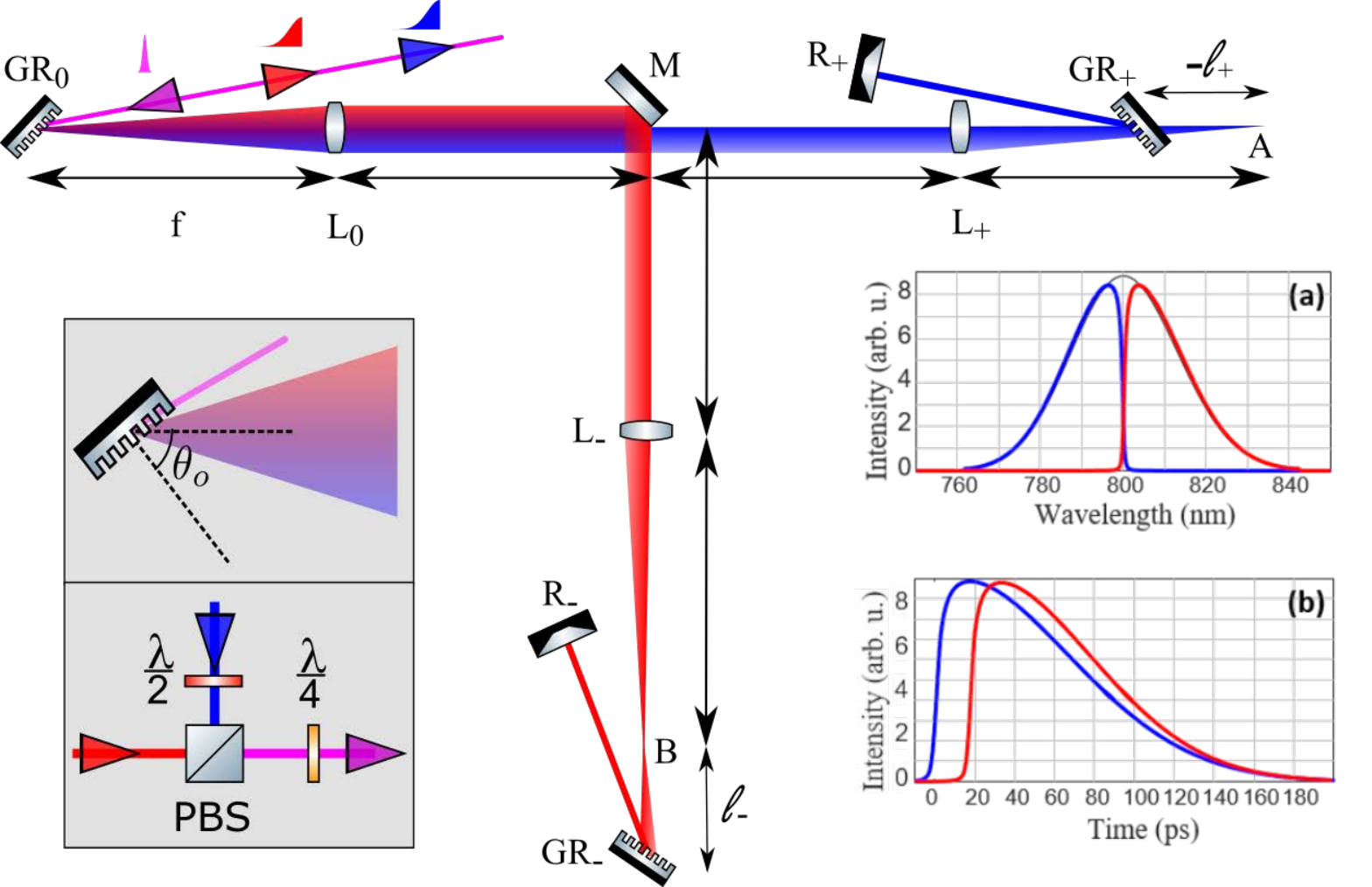}
    \caption{ Centrifuge pulse shaper. Shown are the gratings (GR), lenses (L) and mirrors (M,R) producing two standard stretchers for the two ``arms'' of the centrifuge. The secondary gratings GR$_\pm$ are displaced from the focal planes of the corresponding lenses L$_\pm$ (points A and B) by distances $l_\pm$. Here, subscripts `$\pm$' refer to the blue (up-chirped) and red (down-chirped) centrifuge fields $E_\pm (t)$, respectively. Retro-reflectors R$_\pm$ shift the beams to the parallel planes above/below the plane of the figure and send both arms back towards the corresponding grating GR$_\pm$. After completing a round trip through the shaper, the two output beams (red and blue arrows) exit GR$_0$ in the direction opposite to the input beam (purple arrow), while being vertically separated from one another. Insets on the left show our definition of the grating orientation angle $\theta_0$, as well as the method of combining the two centrifuge arms into a single centrifuge pulse by means of two wave plates and a polarizing beam splitter (PBS). For the fast (slow) centrifuge implemented in our laboratory, the main parameters are as follows: grating groove density $1/d= 1500$ (2400)~mm$^{-1}$, grating orientation angle $\theta_0 \approx 30$ (70)~degrees, effective grating spacing $\left| l_\pm \right| \approx 30$~cm, and focal length $f=20$ (50)~cm. See text and Fig.~\ref{fig:BetaContour} for the detailed discussion of the reason behind choosing these values. Examples of the calculated spectral and temporal profiles of an optical centrifuge are shown in insets \textbf{(a)} and \textbf{(b)} on the right, respectively. Unequal frequency chirps, and hence unequal pulse durations, as well as unequal optical path lengths, and hence mismatched rising edges, were assumed for illustration purposes.}
    \label{fig:CFGshaper}
\end{figure*}

\section{Making an Optical Centrifuge}
\label{sec:MakingCFG}

As originally proposed and implemented by Villeneuve \textit{et al.} \cite{Villeneuve2000}, the field of an optical centrifuge can be created by means of a pulse shaper shown in Figure~\ref{fig:CFGshaper}. The shaper consists of two standard pulse stretchers based on diffraction gratings \cite{Treacy1969}, which produce the two frequency-chirped pulses $E_\pm$ described in Eq.~(\ref{eq:Eplusminus}). The stretchers are built in a well-known `$4f$' geometry \cite{Weiner2011}, in which an image of the input grating (GR$_0$, shared by the two stretchers ), is created by means of a lens pair (L$_0$/L$_+$ and L$_0$/L$_-$) and located four focal lengths away at points A and B. The $4f$ geometry is utilized for two purposes. First, it enables splitting the input beam into two spectral components, hereafter referred to as ``two centrifuge arms'', in the Fourier plane of lens L$_0$. Numerically modeled spectra for each arm of fCFG are shown in inset \textbf{(a)}, whereas inset \textbf{(b)} shows an example of the corresponding temporal profiles with slightly different durations and time delays (with respect to the input pulse) to illustrate the case of a misaligned centrifuge, often found in a real experiment. Second, the $4f$ configuration makes it possible to apply frequency chirps of both positive and negative sign. Following Ref.~\citenum{Treacy1969}, the chirp produced by a standard double-pass double-grating stretcher around the center frequency $\omega_0$ can be calculated as:
\begin{equation}\label{eq:BetaStretcher}
\beta_\pm=-\frac{d^2\omega_0^3\cos^2(\theta_0)}{16\pi^2l_\pm c},
\end{equation}
\noindent where $l_\pm$ is the distance between the gratings along the path of the light beam with $\omega=\omega_0 $, $1/d$ is their groove density, $\theta_0$ is the orientation angle of the gratings (see Figure~\ref{fig:CFGshaper}), and $c$ is the speed of light. To arrive at the above expression, we used the inverse relationship between the second derivatives of the spectral and temporal phases of the field with a gaussian envelope in the limit of large pulse stretching \cite{RulliereBook}, and took into account the double passage of centrifuge pulses through the $4f$ shaper. One can see that the sign of $\beta _\pm$ is determined by the sign of the corresponding $l_\pm$. Imaging an input grating at locations A and B enables negative values for the effective distance between the gratings (here, negative $l_+$ in the blue centrifuge arm in Figure~\ref{fig:CFGshaper}) and hence positive values of the frequency chirp.
\begin{figure*}[t]
    \centering
    \includegraphics[width=.9\textwidth]{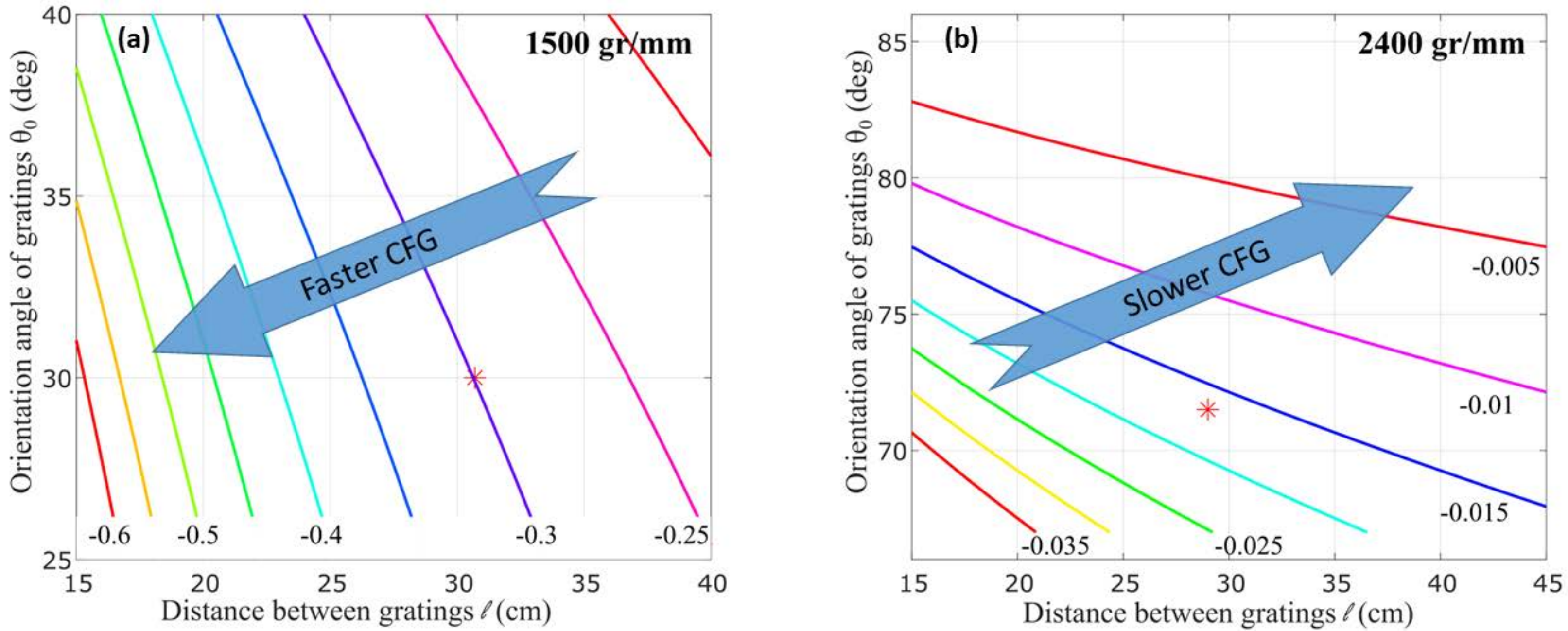}
    \caption{Frequency chirp $\beta $, defined in Eq.~\ref{eq:BetaStretcher}, as a function of the separation distance and angle of the diffraction gratings in the centrifuge shaper for the central wavelength $\lambda_{0}=800$~nm. \textbf{(a)} Fast centrifuge limit and \textbf{(b)} slow centrifuge limit, with gratings of 1500 and 2400 grooves per mm, respectively. Lines are labeled with the corresponding values of $\beta $ in rad/ps$^{2}$. Red asterisks mark the parameters chosen for the two centrifuges implemented in our laboratory.}
    \label{fig:BetaContour}
\end{figure*}

According to Equation~(\ref{eq:BetaStretcher}), to make the centrifuge accelerate faster (higher absolute values of $\beta $), one has to decrease distance $l:=|l_\pm|$, angle $\theta_0$, and groove density $1/d$, as illustrated in Figure~\ref{fig:BetaContour}. From the practical perspective, this does not present any technical challenges since weaker gratings are readily available and shorter distances are easy to implement. If one wishes to use uncompressed pulses from a typical Ti:Sapphire chirped pulse amplifier, as in the original work of Villeneuve \textit{et al.}\cite{Villeneuve2000}, an output frequency chirp on the order of $+0.3$~rad/ps$^{2}$ must be taken into account. In practice, this amounts to shifting the two gratings in both centrifuge arms by the same positive distance, i.e. $l_\pm \rightarrow l_\pm + \Delta l$, dependent on the grating groove density. This dictated the choice of parameters for our fast centrifuge ($\beta =0.3$~rad/ps$^{2}$), marked with a red asterisk in Figure\ref{fig:BetaContour}\textbf{(a)}, whose blue arm is built with $l_+ +\Delta l \approx 0$ for convenience.

To make the centrifuge accelerate slower (lower $\beta$), one needs to increase the three shaper parameters ($l,\theta_0$ and $1/d$), which are all capped by hard technical constraints. The highest available groove density producing reasonably high diffraction efficiency in the spectral region of interest is 2400~mm$^{-1}$. As the angle between the incident beam and the normal to the grating is larger than $\theta_0$ (by the deviation angle of a few degrees), the latter is naturally limited by $\approx 80^{\circ}$. Finally, the absolute value of distance $l_+$ must be smaller than the focal length $f$, or else the diffracted beam in the blue arm will not clear the lens (see Figure~\ref{fig:CFGshaper}). These considerations determined the parameters of our slow centrifuge ($\beta =0.017$~rad/ps$^{2}$) marked in plot \textbf{(b)} of Figure~\ref{fig:BetaContour}.

The spectral bandwidth of the centrifuge shaper $\Delta \nu_{\text{sh}}$, and hence the maximum angular frequency of the centrifuge $\nu_{\text{max}}=\Delta \nu_{\text{sh}} /2$, is determined by the groove density of the gratings and the numerical aperture ${NA}$ of the lenses. Using the grating equation and a limit of small $NA$ (typical for most pulse shapers due to geometrical constraints), one arrives at $\Delta \nu_{\text{sh}} = 2dc\cos(\theta_0){NA}/\lambda^2_0$. Using Equation~(\ref{eq:BetaStretcher}), the terminal frequency of the centrifuge can be expressed as a function of the chirp parameter:
\begin{equation}\label{eq:Bandwidth}
\nu_{\text{max}}={NA} \sqrt{\beta \frac{2l}{\pi\lambda_0}}.
\end{equation}
\noindent For the two centrifuge configurations indicated with red asterisks in Figure~\ref{fig:BetaContour}\textbf{(a,b)}, the maximum achievable rotation frequencies are therefore 22.2 and 3.3~THz. Here, the numerical aperture was calculated using the focal length $f=20$~cm and $f=50$~cm for the fast and slow centrifuge, respectively, as implemented in our experimental setups.

The spectral bandwidth of the laser source should be matched to that of the centrifuge shaper. If the former is too broad, valuable energy will be lost with photons whose frequencies are outside of the shaper's frequency window. In the opposite limit of input pulses which are spectrally too narrow, there will be not enough field bandwidth to accelerate molecular rotation all the way to $\nu_{\text{max}}$. Using the values of $\Delta \nu_{\text{sh}}=2\nu_{\text{max}}$ calculated above for the two suggested sets of centrifuge parameters, one can see that the two centrifuges will offer best functionality when matched with two different sources of high energy femtosecond pulses. The fast centrifuge is best matched with a source of $\approx 30$~fs pulses, whose full bandwidth is typically on the order of 25~THz, whereas a narrow band source of $\approx 130$~fs pulses (full bandwidth of about 6~THz) will better match the slow centrifuge.

\section{Characterization of an optical centrifuge}
\label{sec:Characterization}
As reviewed earlier in the text (Sec.~\ref{sec:Theory}), the field of an optical centrifuge is a result of an interference of two circularly polarized pulses [or ``arms'', $\vec{E}_\pm(t)$], described by Equation~(\ref{eq:Eplusminus}). The two arms are obtained by splitting the initial laser pulse in the Fourier plane of the pulse shaper, as illustrated in Figure~\ref{fig:CFGshaper}. Their numerically calculated spectra are shown in Figure~\ref{fig:CFGshaper}\textbf{(a)}. Applying negative and positive frequency chirp to the red and blue arm, respectively, results in the corresponding temporal envelopes shown in Figure~\ref{fig:CFGshaper}\textbf{(b)}. For illustration purposes, the absolute values of two frequency chirps were assumed slightly different, resulting in a small difference between the duration of two arms. Optimal interference is achieved when the two profiles overlap well in time, which relies on the overlap of their rising edges and equal pulse lengths, with both of those parameters being controlled by the shaper elements and geometry.

Below we describe three different methods of assessing the degree of this temporal overlap and measuring the rotational frequency of the generated centrifuge pulse. All experimental results correspond to the fast and slow optical centrifuges, whose technical parameters are indicated by the corresponding asterisk in Fig.~\ref{fig:BetaContour}. The two centrifuges have been built with two separate laser sources, both generating ultrashort pulses at 1~KHz repetition rate and 800~nm central frequency. Pulse energies and spectral widths at the input of the fast (slow) centrifuge shaper are as follows: 10~mJ (15~mJ) per pulse and 29~nm (7~nm) full width at half maximum.
\begin{figure*}[bt]
    \centering
    \includegraphics[width=.9\textwidth]{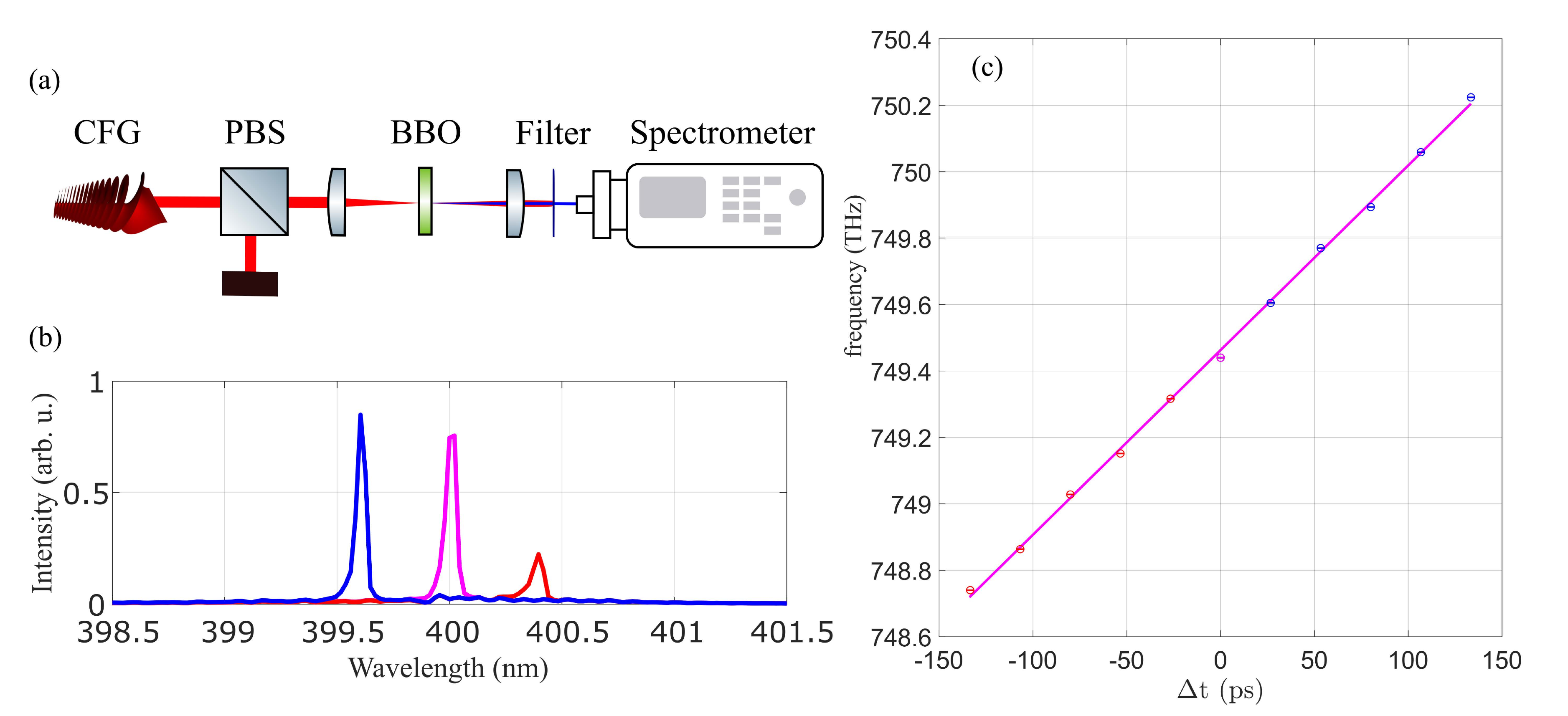}
    \caption{\textbf{(a)} Layout of the experimental setup for the pileup technique of centrifuge characterization. Centrifuge pulses are linearly polarized with a polarizing beam splitter (PBS) and focused on a nonlinear barium borate crystal (BBO). The remaining red light is filtered, and the sum frequency generation (SFG) beam is sent to a spectrometer.  \textbf{(b)} Examples of pileup spectra at different time delays between the two centrifuge arms. The line width of $\approx 0.1$~nm is dictated by the spectrometer resolution. Much broader second harmonic spectra from each individual arm produce a low-intensity pedestal, almost invisible on the scale of the plot. The worse the quality of the centrifuge, the weaker the pileup peak with respect to the pedestal. \textbf{(c)} Measured dependence of the SFG central frequency on the time delay between the two centrifuge arms. The value of $\beta$ determined from the slope of the curve is 0.0175(5)~rad/ps$^{2}$.}
    \label{fig:Pileup}
\end{figure*}

\subsection{Pileup method}
\label{sec:Pileup}
The pileup method is the simplest way to assess the quality of the centrifuge field. It is based on the frequency mixing of the two centrifuge arms on a nonlinear crystal, such as barium borate (BBO), and recording the spectrum of the sum frequency generation (SFG) signal as a function of the relative time delay $\Delta t $ between the two pulses. The required experimental arrangement is shown schematically in Figure~\ref{fig:Pileup}\textbf{(a)}. The centrifuge field is linearly polarized and focused on a nonlinear crystal. The spectrum of the SFG light is recorded as a function of the time delay, controlled by moving one of the retro-reflectors in the centrifuge shaper (R$_\pm$ in Figure~\ref{fig:CFGshaper}).

The SFG pileup field is proportional to the product of the two fields $E_\pm$ making up the centrifuge (all linearly polarized, e.g. along $x$ axis),
\begin{equation}\label{eq:Pileup}
E_{\text{pu}}(t,\Delta t) \propto E_{+,x}(t+\Delta t/2) E_{-,x}(t-\Delta t/2),
\end{equation}
\noindent where $E_{\pm,x}$ are the two arms of the centrifuge and $\Delta t$ is the time delay between them. Using Equation~(\ref{eq:Eplusminus}), the time-dependent part of the phase of the pileup field is $\varphi_{\text{pu}}(t) = \left[ 2\omega _0\ + 2\beta \Delta t \right]t + \Delta \beta t^2$, where we introduced $\Delta \beta \equiv (\beta _+ + \beta _-)$ (remember that $\beta _+>0$ and $\beta _- <0$). The pileup pulse is therefore a frequency chirped pulse with a time-dependent instantaneous frequency
\begin{equation}\label{eq:PileupFrequency}
\omega _{\text{pu}}(t) = 2\omega _0 + 2\beta \Delta t +2 \Delta \beta t.
\end{equation}
\noindent In the case of equally chirped arms, $\Delta \beta =0$, the chirp disappears and the spectrum ``piles up'' at a single frequency of $2\omega _0 + 2\beta \Delta t$ (similarly to the spectral narrowing of the second harmonic of a pulse with an antisymmetric phase \cite{Comstock2004}). Examples of experimentally recorded SFG spectra at different values of time delay $\Delta t$ are shown in Figure~\ref{fig:Pileup}\textbf{(b)} for the case of our slow centrifuge. Strong narrow peaks on top of a weak broad pedestal (due to the second harmonic of each individual centrifuge arm) are indicative of a pileup effect. Uneven chirping, i.e. $\Delta \beta \neq 0$, leads to the spectral broadening of the pileup peak, therefore  allowing a convenient way of equalizing the two chirp rates by adjusting the grating positions $l_{\pm}$ in the centrifuge shaper for the narrowest pileup spectrum. Equal frequency chirps of the two centrifuge arms result in their equal pulse lengths and, correspondingly, the most optimal overlap in time and the highest quality of the centrifuge.

Equation (\ref{eq:PileupFrequency}) shows that the pileup frequency is linearly proportional to the time delay between the centrifuge arms. The observed linear dependence is plotted in Figure~\ref{fig:Pileup}\textbf{(c)}. From the detected slope of 5.6~GHz/ps, one can extract the average frequency chirp parameter $\beta\equiv (\beta _+ - \beta _-)/2 = 0.0175(5)$~rad/ps$^{2}$.
\begin{figure*}[t]
    \centering
    \includegraphics[width=.9\textwidth]{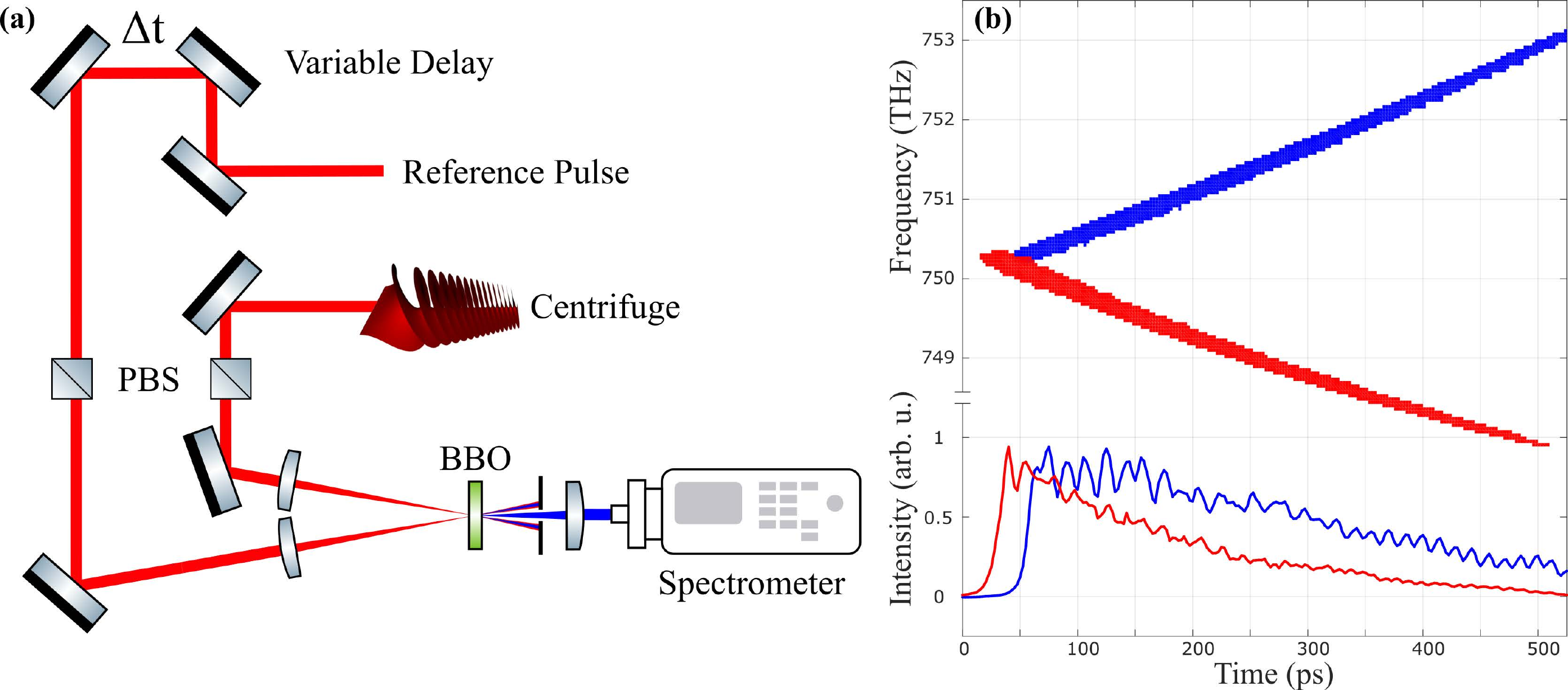}
    \caption{ \textbf{(a)} Experimental layout for the cross-correlation frequency resolved optical gating (XFROG) characterization of the centrifuge field. Centrifuge pulses are linearly polarized with a polarizing beam splitter (PBS) and focused on nonlinear barium borate (BBO) crystal, together with the linearly polarized short reference pulses. Sum frequency generation spectrum is recorded with a spectrometer as a function of the time delay $\Delta t$. \textbf{(b)} top: XFROG scans of the red (tilted up) and blue (tilted down) centrifuge arms of a slow centrifuge; bottom: temporal profiles obtained from the traces above. The nonzero time delay between the rising edges can be easily eliminated and was left here for illustration purposes [cf. Fig.~\ref{fig:CFGshaper}\textbf{(b)}].}
    \label{fig:XFROG}
\end{figure*}

\subsection{XFROG}
Cross-correlation frequency-resolved optical gating (XFROG) is a well known method of characterizing laser pulses \cite{Linden1998,Linden1999}, schematically shown in Figure~\ref{fig:XFROG}\textbf{(a)}. Similarly to the pileup technique described above, the method is based on sum frequency generation. Here, however, the field of each centrifuge arm is mixed with a short Fourier-transform limited reference pulse. The spectrum of the XFROG field,
\begin{equation}\label{eq:XFROG}
E_{\text{XFROG},\pm}(t,\Delta t) \propto E_{\pm,x}(t+\Delta t/2) E_{\text{TL},x}(t-\Delta t/2),
\end{equation}
\noindent is again recorded as a function of the time delay $\Delta t$. Reference field $E_{\text{TL}} (t)$ corresponds to a compressed pulse, split from the main laser output before the centrifuge shaper. As $E_{\text{TL}} (t)$ is much shorter than the centrifuge pulse ($\sim1$~ps \textit{vs} $>100$~ps, respectively), the recorded XFROG spectrograms $\left|E_{\text{XFROG},\pm}(\omega ,\Delta t )\right|^2$ are direct time-frequency maps of each centrifuge arm. They reflect the time dependence of the instantaneous frequency of $E_\pm(t)$, shifted to $2\omega _0$ due to the SFG mixing, i.e. $2\omega _0 + 2\beta_\pm \Delta t$. Hence, the slope of each XFROG spectrogram corresponds to twice the frequency chirp of the corresponding arm, $\beta_\pm$.

An example of the experimentally measured XFROG trace for our slow centrifuge is shown in the upper part of Figure~\ref{fig:XFROG}\textbf{(b)}. Unlike the pileup technique, XFROG allows to assess the quality of each arm separately. For instance, the plotted traces reveal a small amount of third order dispersion (TOD), visible through the slight curvature in each trace. The latter is expected to have the same sign for both centrifuge arms, as it is proportional to $\left(\beta _{\pm} \right)^2$ (Ref.~\citenum{Treacy1969}). In addition, integrating over wavelength provides one with a temporal profile of each arm, shown in the lower part of Figure~\ref{fig:XFROG}\textbf{(b)}. In contrast to the harmless TOD curvature, a clearly visible mismatch between the two rising edges, as well as between the decaying tails of each arm, will affect the performance of the centrifuge as they will result in a non-zero initial and lower terminal frequency of the rotating polarization, respectively. Fortunately, both defects can be easily eliminated by adjusting the positions of mirror M and retro-reflectors R$_\pm$ (see Fig.~\ref{fig:CFGshaper}).

Intensity oscillations (especially in the blue arm) may also be detrimental to the spinning efficiency. They stem from the oscillations in the spectrum of the initial laser pulse and should be minimized by the proper alignment of the laser cavity. Similar alignment should also be executed to equalize the intensity profiles of the two arms, seen in the bottom plot of Fig.~\ref{fig:XFROG}\textbf{(b)}. Unequal intensities will result in a small degree of ellipticity of the centrifuge polarization and lower its spinning performance.

From the two initial slopes of $-5.4(2)$~GHz/ps and $5.8(2)$~GHz/ps, one can extract the frequency chirps of $\beta _{+}=0.0171(5)$~rad/ps$^{2}$ and $\beta _{-}=-0.0182(5)$~rad/ps$^{2}$, for the blue and red arms, respectively. The average value of $\beta =0.0176(7)$~rad/ps$^{2}$ is in good agreement with the pileup method described in Sec.~\ref{sec:Pileup}.
\begin{figure}[b!]
    \centering
    \includegraphics[width=.8\columnwidth]{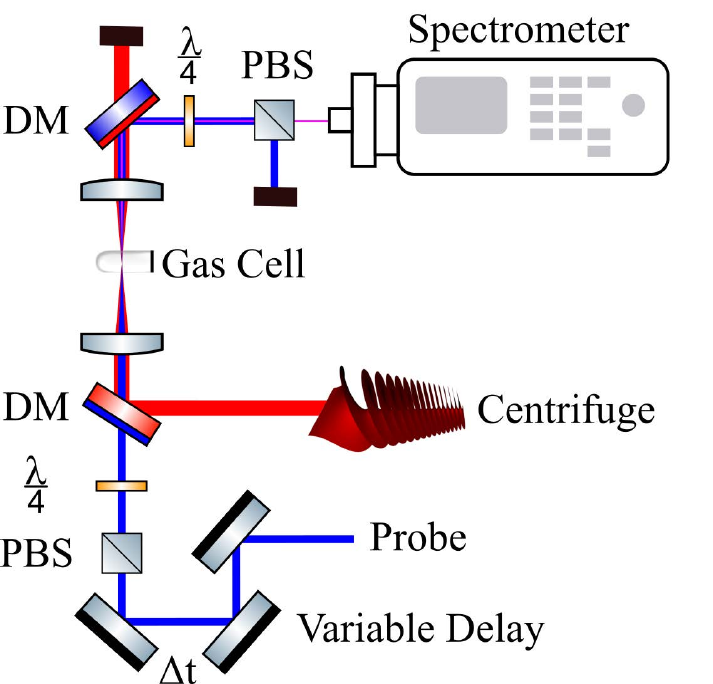}
    \caption{Experimental configuration for characterizing the rotational excitation with an optical centrifuge by means of coherent Raman spectroscopy. PBS: polarizing beam splitter, DM: dichroic mirror. An optical centrifuge field is produced in the centrifuge shaper shown in Figure~\ref{fig:CFGshaper}. Probe pulses are compressed pulses, split from the main laser output before the centrifuge shaper and frequency doubled in a nonlinear crystal.}
    \label{fig:Raman_setup}
\end{figure}

\subsection{Raman Spectroscopy}
Coherent Raman scattering (CRS) of light from rotating molecules represents one of the most direct detection techniques to characterize the rotational excitation by an optical centrifuge. It enables one to measure both the frequency and the direction of molecular rotation by recording the frequency shift and the polarization of the scattered light, respectively. Applying CRS to centrifuged molecules has been described in a number of our previous publications (for a review, see Ref.~\citenum{Milner2016a}). Briefly, owing to the centrifuge-induced coherence between the quantum states separated by $\Delta J=\pm2$ (with $J$ being the rotational quantum number), the spectrum of a probe light passing through the centrifuged ensemble develops Raman sidebands. The magnitude of the Raman shift equals twice the rotation frequency, while its sign reflects the direction of molecular rotation with respect to the probe's circular polarization: Stokes down-shifting indicates molecular rotation in the same direction as the probe polarization, whereas anti-Stokes up-shifting means the directions are opposite to one another. The polarization of the Raman sidebands is circular and opposite to the input probe polarization.
\begin{figure*}[tb]
    \centering
    \includegraphics[width=0.9\textwidth]{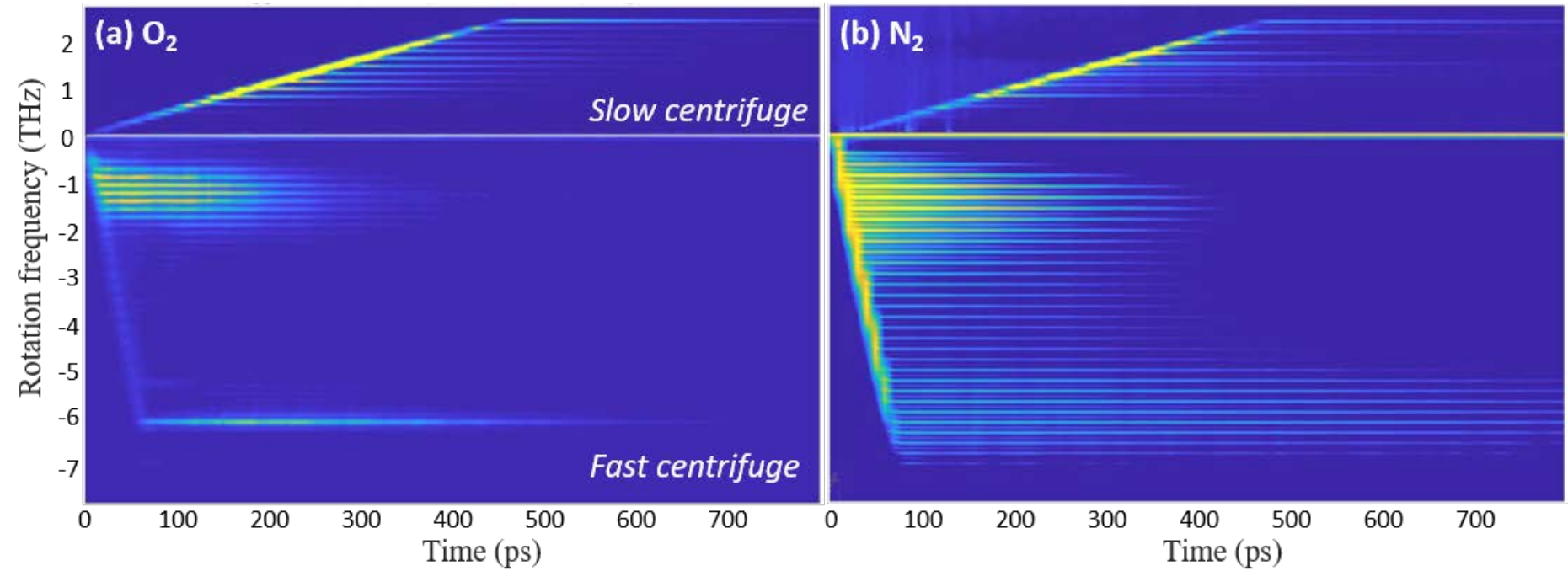}
    \caption{Two-dimensional Raman spectrograms from the centrifuged gas of \textbf{(a)} oxygen and \textbf{(b)} nitrogen. Positive and negative frequency shifts correspond to the data taken (separately) with a slow and fast centrifuge, respectively, and combined into a single spectrogram for comparison. Horizontal traces reflect the centrifuge-induced rotational coherence, which decays due to collisions. Tilted traces starting from the origin are due to the linearly increasing with time rotational frequency of centrifuged molecules.}
    \label{fig:Raman}
\end{figure*}

The Raman setup is shown schematically in Figure~\ref{fig:Raman_setup}. Centrifuge pulses from the centrifuge pulse shaper, described in Sec.~\ref{sec:MakingCFG}, are focused inside an optical cell filled with a gas at room temperature and variable pressure. Probe pulses, circularly polarized by a combination of a linear polarizer and a quarter-wave plate, follow the centrifuge after a variable time delay. Frequency doubling of probe pulses allows us to combine and separate them from the centrifuge light using dichroic mirrors. Another set of a linear polarizer and a wave plate acts as a circular analyzer, separating the weak frequency shifted Raman sidebands from the strong unshifted and oppositely polarized background. Narrowing the spectral bandwidth of probe pulses to $\approx0.1$~nm by means of the probe pulse shaper is useful for resolving individual rotational states, as well as for distinguishing centrifuge-populated superrotor states from those initially populated in a room-temperature ensemble.

Examples of CRS spectra from the centrifuged gas of oxygen and nitrogen molecules are shown in Figure~\ref{fig:Raman}\textbf{(a)} and \textbf{(b)}, respectively. Recording the spectra as a function of the centrifuge-probe time delay provides us with a two-dimensional Raman spectrogram and offers a convenient way of following the rotational dynamics of molecules inside the centrifuge, as well as the field-free dynamics at later times. For both gases, Raman spectrograms recorded with a slow centrifuge (upper traces) and a fast centrifuge (lower traces) were taken in separate optical setups and different gas cells. They are plotted against one another for the purpose of comparing the action of the two centrifuge types.

The length of the slow centrifuge is about 500~ps. During this time, the molecules are following its accelerated rotation up to about 2.5~THz, as reflected by the bright tilted traces in the upper part of the spectrograms. From the tilt, one can extract the angular acceleration (5.4~GHz/ps) and the corresponding value of the average frequency chirp $\beta _{\text{sCFG}}=0.0170(3)~$rad/ps$^{2}$, in agreement with the chirp values extracted from the pileup and XFROG methods. Some molecules fall behind the centrifuge, while others are brought to the superrotor states at the end of the sCFG pulse. In the case of the fast centrifuge, the molecules spin up to about 7~THz in about 70~ps (95.5~GHz/ps), corresponding to the average frequency chirp $\beta _{\text{fCFG}}=0.31(1)$~rad/ps$^{2}$.

\section{Conclusion}
\label{sec:Conclusion}
In summary, we have reviewed the utility of an optical centrifuge as a powerful tool for controlling molecular rotation in a wide range of physical parameters. We have discussed how the key properties of the centrifuge -- its optical intensity $I$ and the rate of its angular acceleration $\beta $, together with the relevant properties of the molecule -- its moment of inertia $\II$ and polarizability anisotropy $\Delta \alpha $, as well as the temperature of the molecular ensemble, can be combined into a dimensionless spinnability factor $\sigma $. By numerically simulating the centrifuge action in the gas of light (oxygen) and heavier (propylene oxide) molecules in two different temperature regimes, thus varying $\sigma $ by two orders of magnitude, we demonstrated the way in which $\sigma $ reflects the likelihood of a molecule to be caught and spun by the centrifuge. We have analyzed quantitatively the amount of centrifuged molecules as a function of the spinnability factor, pointing out the advantage of slowing the centrifuge rotation for spinning molecules with lower $\Delta \alpha/\II $ ratio.

Technical challenges of building an optical centrifuge with either high or low angular acceleration have been discussed in detail. We have outlined realistic design parameters for both acceleration limits, referred to as ``fast'' and ``slow'' centrifuge. We have shown that slowing the centrifuge by a factor of almost 20 (in terms of the value of $\beta $) is feasible, but comes at the expense of lower (by a factor of 4) terminal rotation frequency.

By building both fast and slow optical centrifuges in our laboratory, we were able to investigate their properties and performances experimentally. Three different characterization techniques have been demonstrated. Two of them -- the ``pileup'' and the XFROG techniques, are based on nonlinear frequency mixing and enable one to measure the centrifuge acceleration rate $\beta$ at relatively low intensity levels, with no requirement to apply the centrifuge in a gas of molecules. Both techniques offer convenient ways of gauging the quality of the centrifuge during the construction phase. The third approach relies on coherent Raman scattering in the gas of centrifuged molecules and provides the direct evidence of the working centrifuge, as well as a detection method of its angular acceleration. We have implemented all three characterization techniques and compared them with one another, providing the examples of experimental observables in each case.

As a well developed laser instrument, an optical centrifuge stimulated a number of recent experimental works and theoretical proposals. Understanding the practical limitations of this tool and the ways of assessing its performance, as described in this work, will help expanding the boundaries of the fascinating field of laser control of molecular dynamics.

\begin{acknowledgments}
This work was carried out under the auspices of the Canadian Center for Chirality Research on Origins and Separation (CHIROS). The authors would like to thank Tsafrir Armon and Lazar Friedland for valuable comments on the topic of parametrizing the efficiency of an optical centrifuge.
\end{acknowledgments}


%

\end{document}